\documentclass[conference, 10pt, a4paper]{IEEEtran}

\newlength{\customcolwidth}
\divide\customcolwidth by 2

\IEEEoverridecommandlockouts
\usepackage{amsfonts}
\usepackage{amssymb}
\usepackage{amsmath}
\usepackage{amsthm}
\usepackage{latexsym}
\usepackage{graphicx}
\usepackage{comment}
\usepackage{tabularray}
\usepackage[table,xcdraw]{xcolor}
\usepackage{algorithm}
\usepackage{algpseudocode}

\usepackage{textcomp}
\usepackage{url}
\usepackage{fancyhdr}
\usepackage[ hidelinks]{hyperref}
\usepackage[utf8]{inputenx}
\usepackage[T1]{fontenc}
\usepackage{amsthm, amsmath, amsfonts}
\usepackage{graphicx}
\usepackage{textcomp}
\usepackage{color, colortbl} 
\usepackage{multirow}
\usepackage[inline]{enumitem}
\usepackage{multicol}
\usepackage[altpo]{backnaur}
\usepackage{booktabs} 
\usepackage{listings}
\usepackage{tikz}
\usepackage{float}
\usepackage{url}
\usepackage{hyperref}
\usepackage{mathpartir}
\usepackage{fancyhdr}
\usepackage{pifont}
\usepackage{xspace}
\usepackage[capitalize]{cleveref}
\usepackage{caption}
\usepackage{array}
\usepackage{boldline}
\usepackage{soul}
\usepackage{listings}
\usepackage{balance}

\usepackage[a4paper, total={184mm,239mm}]{geometry}
\def\BibTeX{{\rm B\kern-.05em{\sc i\kern-.025em b}\kern-.08em
    T\kern-.1667em\lower.7ex\hbox{E}\kern-.125emX}}

\newcommand{\secmc}{SecIC3\xspace}

\algrenewcommand\algorithmiccomment[1]{\hfill{\footnotesize$\triangleright$~#1}}

\newcommand{\proc}[1]{\textsc{#1}}

\newcommand{\eg}{e.g.,\xspace}

\newif\ifdraft
\drafttrue  
\draftfalse

\ifdraft
  
  \newcommand{\qinhan}[1]{\textsf{\footnotesize\color{blue}[Qinhan: #1]}}
  \newcommand{\yuwei}[1]{\textsf{\footnotesize\color{teal}[Yu-Wei: #1]}}
  \newcommand{\akash}[1]{{\color{orange}#1}}
  \newcommand{\akashsays}[1]{\textsf{\footnotesize\akash{[Akash: #1]}}}
  
\fi

\begin{document}
\title{SecIC3: Customizing IC3 for\\ Hardware Security Verification}


\author{%
Qinhan Tan, Akash Gaonkar, Yu-Wei Fan, Aarti Gupta, and Sharad Malik \\
\textit{Princeton University, Princeton, New Jersey, USA} \\
\{qinhant, agaonkar, yf9172, aartig, sharad\}@princeton.edu
}


\maketitle

\let\spacysection\section
\renewcommand{\section}[1]{\vspace{-0.4em}\spacysection{#1}\vspace{-0.2em}}
\let\spacysubsection\subsection
\renewcommand{\subsection}[1]{\vspace{-0.4em}\spacysubsection{#1}\vspace{-0.2em}}

\begin{abstract}
Recent years have seen significant advances in using formal verification to check hardware security properties. 
Of particular practical interest are checking confidentiality and integrity of secrets, by checking that there is no information flow between the secrets and observable outputs. 
A standard method for checking information flow is to translate the corresponding non-interference hyperproperty into a safety property on a \emph{self-composition} of the design, which has two copies of the design composed together. 
Although prior efforts have aimed to reduce the size of the self-composed design, there are no state-of-the-art model checkers that exploit their special structure for hardware security verification.
In this paper, we propose \secmc, a hardware model checking algorithm based on IC3 that is customized to exploit this self-composition structure.
\secmc utilizes this structure in two complementary techniques: \textit{symmetric state exploration} and \textit{adding equivalence predicates}.
We implement \secmc on top of two open-source IC3 implementations and evaluate it on a non-interference checking benchmark consisting of 10 designs.
The experiment results show that \secmc significantly reduces the time for finding security proofs, with up to 49.3x proof speedup compared to baseline implementations.

\end{abstract}

\begin{IEEEkeywords}
Hardware Security, Non-Interference Property, Model Checking
\end{IEEEkeywords}

\section{Introduction}
\label{sec:intro}
Hardware security has received increasing attention in recent years as malicious attackers have used hardware vulnerabilities to compromise secret information. Breaches to \emph{confidentiality}, where attackers steal secrets, and \emph{integrity}, where attackers modify them, can result in economic loss~\cite{IBMreport} and threaten public safety~\cite{shield2015hardware}.
%
%
%
One defensive approach that provides strong hardware security guarantees is \emph{formal hardware security verification}, which uses formal verification techniques to systematically check a hardware implementation against a given attack model.
%
Formal verification uniquely provides \emph{exhaustive} security guarantees by constructing a rigorous 
security proof reasoning over all possible executions of a hardware design, and potentially identifying vulnerabilities that may have been missed by other (simulation-based) approaches, such as fuzzing~\cite{hur2022specdoctor,cellift, xu2025dejavuzz}.
However, current work in formal security verification~\cite{IODINE,tan2025rtl,yang2023pensieve,wang2023specification} has been limited to small hardware systems and does not scale to real-world designs.
In this paper, we propose to address this scaling gap by customizing the  model checking algorithm commonly used for proving security properties, particularly confidentiality and integrity.


\begin{figure}[t]
    \centering
    \includegraphics[width=0.6\columnwidth]{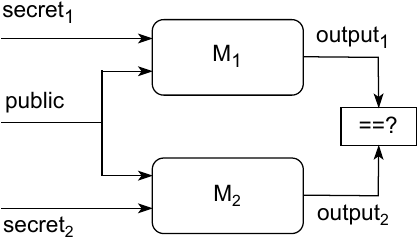}
    \caption{Non-Interference Checking via Self-Composition. 
    }
    \vspace{-1.5em}
    \label{fig:selfcomposition}
\end{figure}

In the literature~\cite{goguen1982security}, both confidentiality and integrity specify the absence of information flow: for confidentiality, changing a secret value should not change any public output, and for integrity, changing a public input should not change any protected data.
More formally, both are forms of a hyperproperty called \emph{non-interference}, in which every pair of executions that differ only in the values of given source variables produce the same values of given sink variables.
We use the terminology of confidentiality throughout the paper. 
%
Seminal work by Barthe et al.~\cite{barthe2004secure} showed that the non-interference hyperproperty can be verified by constructing a \emph{self-composition}, as depicted in \cref{fig:selfcomposition}.
Two copies of a system are composed together, such that only the secret inputs are allowed to differ between the two copies. This simulates a pair of executions in parallel, so that one can prove non-interference by showing that corresponding outputs between the two copies are always equal --- a safety property over the self-composition.

To our knowledge, prior work in formal hardware security verification~\cite{tan2025rtl,tan2023security} has discharged the non-interference verification condition using generic model checking algorithms~\cite{bradley2011sat}, treating the self-composition as a black-box transition system without taking advantage of its symmetric structure. 
Beyond security verification, previous work in general model checking (e.g., ~\cite{ESCav93,tang2005symmetry}) has benefited from symmetry reduction. These were based on 
traditional model checking algorithms, and while their techniques do not translate directly to our context, are inspirational to our work. 

This paper proposes \secmc, a customization of the state-of-the-art IC3 model checking algorithm~\cite{bradley2011sat,een2011efficient} to make it more effective at verifying non-interference.
As part of \secmc we introduce two complementary techniques that improve solver performance by exploiting the special structure of the non-interference self-composition:

\paragraph{Exploring symmetric state}
Since a self-composition consists of two identical copies of the same design, its space of reachable states is highly symmetric. 
In particular, a copy-relabeling mapping allows us to take any reachable state, swap corresponding variables of the two copies, and produce another reachable state.
We augment the model checker with this relabeling mapping and use it to strengthen the blocking step of IC3 by learning additional lemmas via relabeling.

\paragraph{Adding equivalence predicates}
For large designs, existing model checkers may struggle to decompose the proof objective relating corresponding outputs of the two copies into subtasks relating corresponding internal subcomponents/variables as this correspondence is not made explicit.
We aid this decomposition by extending the self-composition with auxiliary \emph{equivalence predicates} that indicate whether corresponding subcomponents between the two copies will output the same value. We modify the IC3 algorithm to better utilize these equivalence predicates, proposing three heuristics for determining which predicates to use during lemma learning, each with their own tradeoffs.


We have implemented two prototypes of \secmc, 
extending the following two open-source IC3 implementations.
\begin{itemize}
\item ABC-PDR~\cite{een2011efficient}: a classic IC3 implementation widely used as a baseline in model checking~\cite{hu2024deepic3, goel2019model, cimatti2016infinite};
\item rIC3~\cite{su2025ric3}: the winner of Hardware Model Checking Competition 2025 on both bit-level and word-level tracks.
\end{itemize}

We also create a new benchmark for non-interference checking, consisting of 10 different open-source hardware designs.
Our evaluations on the benchmark show that
\secmc can improve the performance of finding security proofs by up to 49.3x times compared to the baseline implementations.
\bigskip

\vspace{-1em}
\noindent\textbf{Summary of Our Contributions}

\begin{itemize}
\item We propose \secmc, an IC3-based model checking algorithm customized for verifying non-interference. 
In the algorithm, we introduce two new techniques, \textit{exploring symmetric state} and \textit{adding equivalence predicates}, that use symmetry within a non-interference verification problem to guide lemma learning within the model checker.
\item We construct a benchmark suite comprising 10 diverse hardware designs specifically for evaluating non-interference checkers.
\item We implement and evaluate \secmc on two open-source IC3 implementations, demonstrating that \secmc achieves up to 49.3x times speedup over baseline implementations in proving non-interference for this benchmark suite.
\end{itemize}

\section{Background: IC3/PDR}
\label{sec:background}

IC3~\cite{bradley2011sat}, also known as \emph{Property Directed Reachability (PDR)}~\cite{een2011efficient} is a state-of-the-art hardware model checking algorithm for unbounded proofs of safety properties.
We will assume some familiarity with the IC3 algorithm, but briefly review the parts relevant to our work. A more complete exposition can be found in Een et al.~\cite{een2011efficient}.

\begin{figure}
    \includegraphics[width=\columnwidth]{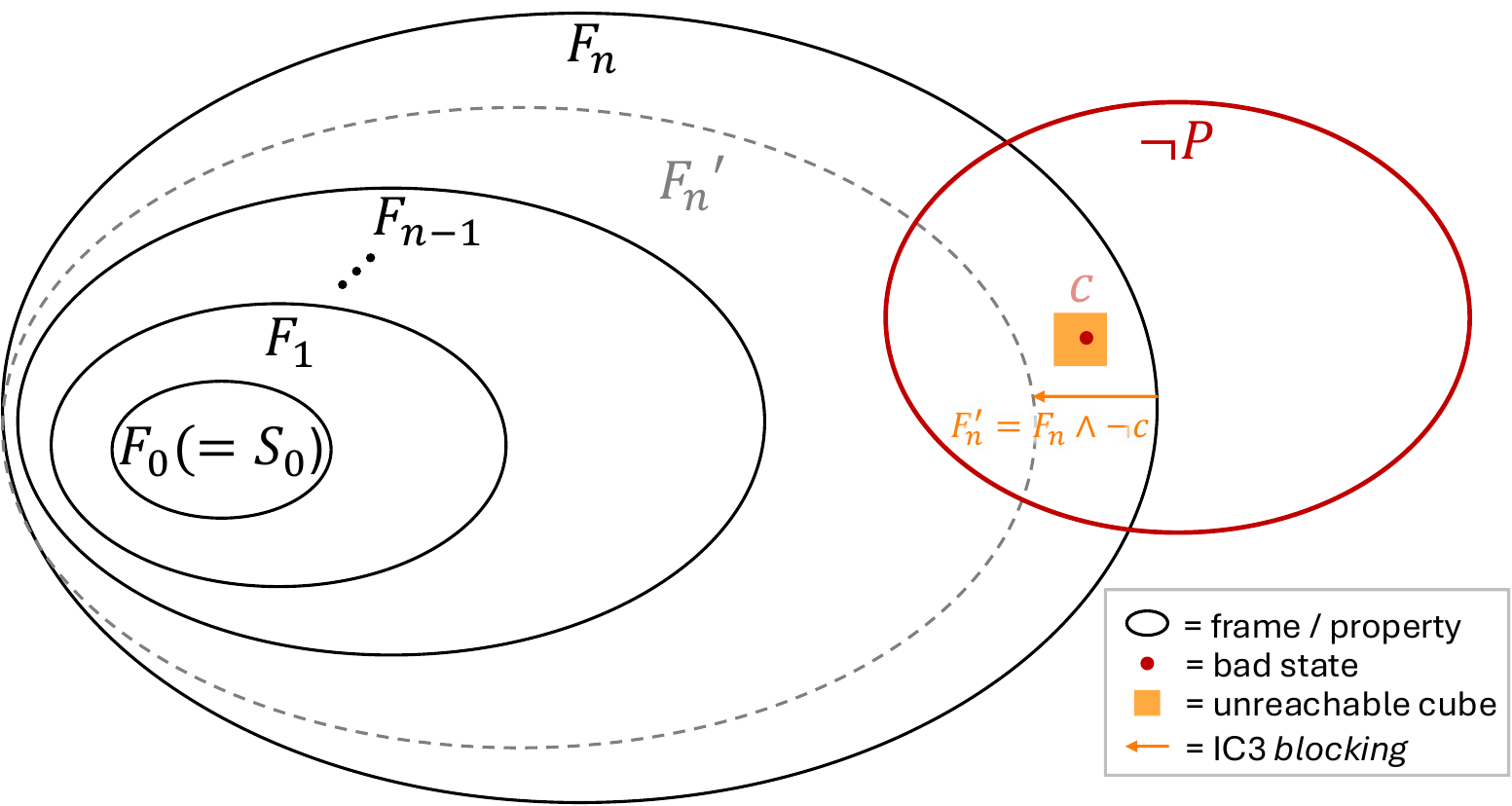}
    \caption{The frames of an IC3 inductive trace when checking a property $P$. IC3 refines frames by \emph{blocking} bad states: the state is shown to be part of an unreachable cube $c$, and is thereby excluded from the latest frame $F_n' = F_n \wedge \neg c$.}
    \vspace{-0.5em}
    \label{fig:ic3}
\end{figure}

Given a transition system $(Init, Tr)$ with state variables $x$, initial states $Init(x)$, and transition relation $Tr(x, x')$, suppose we wish to check a safety property $P(x)$. 
(We follow the standard notation of using a variable $x$ to refer to current state, and $x'$ as the next state.)
As depicted in \cref{fig:ic3},
IC3 tries to construct a safe inductive invariant proving $P$ by maintaining an \emph{inductive trace}, a sequence of formulas, $F_0, F_1, \ldots, F_n$, called \emph{frames}.
Each frame $F_k$ represents an over-approximation of the reachable states within $k$ transitions, with all but the latest frame proven safe with respect to $P$. Formally,
\begin{enumerate}
    \item $F_0(x) = Init(x)$
    \item $\forall k < n, F_k(x) \implies F_{k+1}(x) $
    \item $\forall k < n, F_k(x) \land Tr(x, x')
    \implies F_{k+1}(x')$ 
    \item $\forall k < n, F_k(x) \implies P(x)$ 
\end{enumerate}



IC3 incrementally refines and adds frames until either the latest frame becomes inductive ($F_n \iff F_{n-1}$) 
or a bad state that violates $P$ is found to be reachable from $Init$.
The former case completes the safety proof, where $F_n$ 
is an inductive invariant that holds on all reachable states of $(Init, Tr)$ 
and that implies $P(x)$.
In the latter case, an execution trace from an initial state to the bad state is returned as a counterexample to the safety of $P$. 

If the current frame is not yet safe for $P$, IC3 needs to refine frames during verification. 
It identifies bad states or their predecessors that can be shown to be unreachable from the previous frame. As shown in \cref{fig:ic3}, an unreachable bad state is generalized into a conjunction of literals (a variable or its negation), called a \emph{cube}. This cube is then \emph{blocked} in frames where it is provably unreachable by updating the frame's formula to include its negation. In the figure, $F_n' = F_n \wedge \neg c$ updates $F_n$ to incorporate the fact that cube $c$ is unreachable within the first $n$ transitions.
This refinement procedure, called \proc{Block}, is the key step studied in our paper.

\algtext*{EndIf}     
\algtext*{EndWhile}  
\algtext*{EndFor}    
\newcommand{\added}[1]{\textcolor{blue}{#1}}
\begin{algorithm}[t]
\caption{\textsc{Block}$(s, k, Tr)$}
\label{alg:block}
\begin{algorithmic}[1]
\Require cube $s$, frame number $k$, and transition relation $Tr$
\If{$k = 0$}
  \State \Return \textbf{false} \Comment{reaches $F_0 = Init$: real counterexample}
  \label{ln:real_cex}
\EndIf
\If{$\mathsf{UNSAT}(F_{k-1} \land \lnot s(x) \land Tr \land  s(x'))$}
\Comment{reachability test}
\label{ln:relative_induction}
  \State $c \gets$ \Call{Generalize}{$ s, k$} \Comment{a generalized cube}
  \label{ln:generalize}
  \For{$j \gets 1$ \textbf{to} $k$}
    \State $F_j \gets F_j \land \lnot c$ 
    \label{ln:strengthen}
  \EndFor
  \State \Return \textbf{true}
  \label{ln:return_block}
\Else
  \State $t \gets$ predecessor cube for $s$ in frame $k-1$
  \If{\Call{Block}{$t,\,k{-}1,\,Tr$}}
  \label{ln:recursive}
    \State \Return \Call{Block}{$s,\,k,\,Tr$}
    \Comment{retry $s$ after predecessor}
    \label{ln:rec_resume_frame}
  \Else
    \State \Return \textbf{false}
  \EndIf
\EndIf
\end{algorithmic}
\end{algorithm}

\cref{alg:block} details how \proc{Block} checks the reachability of and potentially blocks a cube $s$ on frame $k$.
\cref{ln:relative_induction} uses a SAT query to check if $s$ is unreachable in frame $k$.
This step is referred to as a \textit{reachability test} in the rest of this paper.
If $s$ is unreachable (UNSAT), a \proc{Generalize} procedure expands $s$ into a cube $c$ describing an unreachable superset of states (essentially by removing literals from $s$).
Cube $c$ is then blocked in \cref{ln:strengthen} by adding $\lnot c$ to frames up to $k$.
On the other hand, if $s$ is reachable (SAT), a predecessor cube $t$ is given by the SAT solver.
IC3 recursively tries to block $t$ in the previous frame $k-1$ (\cref{ln:recursive}), returning to the current frame if successful (\cref{ln:rec_resume_frame}) or exiting with a failing execution trace if frame 0 is reached (\cref{ln:real_cex}).

\section{\secmc Design}
\label{sec:design}
In this section, we detail our customizations to IC3 that use the symmetry of a self-composed design to improve performance when verifying non-interference properties.

\subsection{Overview}

\begin{figure}[t]
    \centering
    \vspace{-0.5em}
    \includegraphics[width=0.8\columnwidth]{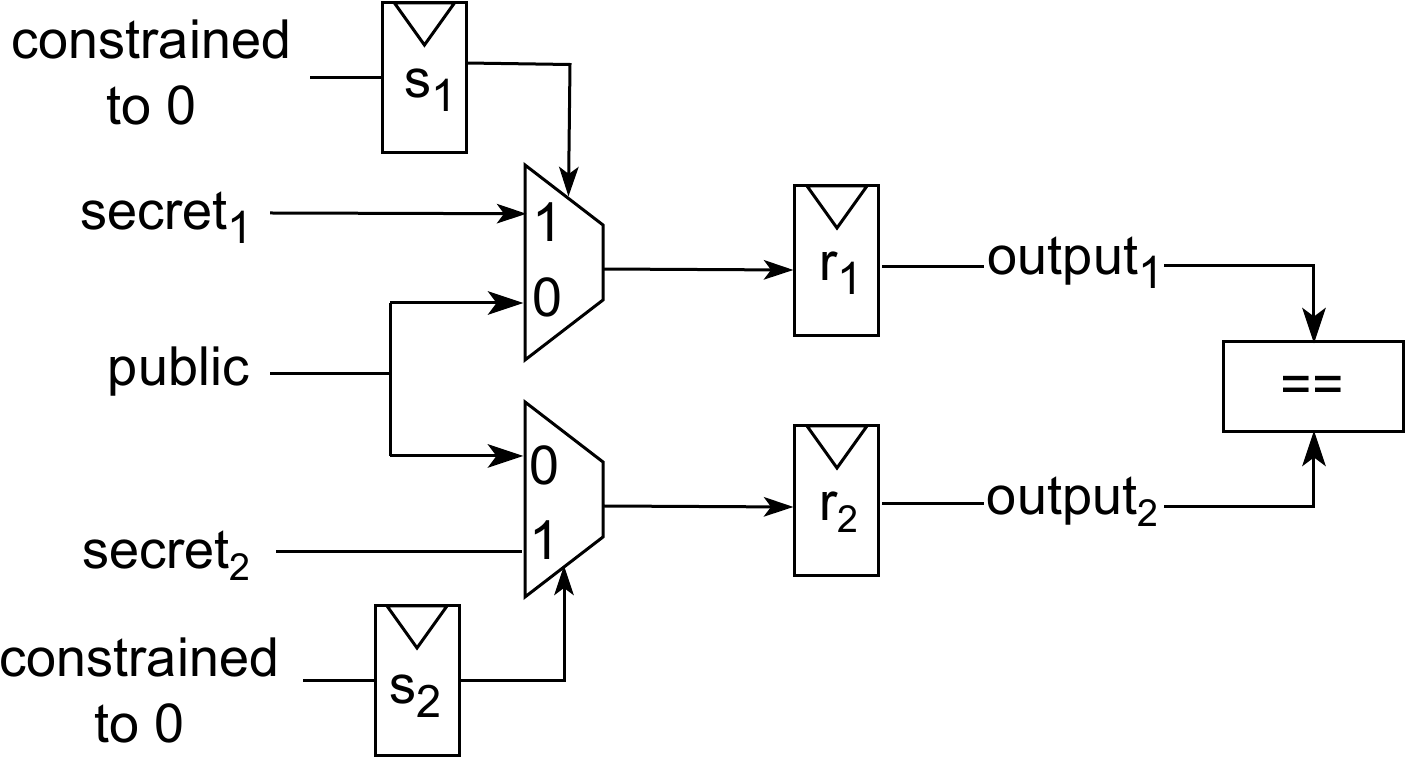}
    
    \caption{IC3 Example on Checking Non-Interference }
    \vspace{-2em}
    \label{fig:motivating_example}
\end{figure}

As an illustrative example, consider the self-composition shown in \cref{fig:motivating_example}, used to check non-interference in a circuit containing a multiplexer and two registers. Call the registers $s$, for selector (1-bit), and $r$, for result ($n$-bit, for some $n$), and suppose also that some input assumptions or combinational logic ensure $s = 0$ always. 
We will use this example to point out some inefficiencies in the traditional IC3 algorithm that motivate our symmetry-based techniques.

When we use IC3 to prove $\text{output}_1 == \text{output}_2$, the algorithm repeatedly calls \proc{Block} on various cubes, one at a time, in order to construct a safe inductive invariant.



\noindent\textbf{Inefficiency 1 (Repeated reasoning).}
For our example, one cube IC3 must block is $(s_1)$, because if $s_1$ is high then in the coming cycles $\text{secret}_1$ will leak into $\text{output}_1$ producing a bad state.
By symmetry, the same logic applies to cube $(s_2)$, but IC3 will likely separately repeat a very similar sequence of bad states and sat queries in order to block this cube also.

\noindent\textbf{Inefficiency 2 (Equalities are expensive).}
After blocking the cubes $(s_1)$ and $(s_2)$, IC3 must prove $r_1==r_2$ always holds by showing that $r_1\neq r_2$ is unreachable.
Crucially, a cube (a conjunction of boolean literals) cannot individually capture complex relational constraints, such as register (in-)equality.
To block all states where the $n$-bit registers $r_1, r_2$ are inequal, IC3 must block two cubes for each bit $i \in [0, n)$:
 $(r_1[i] \land \lnot r_2[i])$ and $(\lnot r_1[i] \land r_2[i])$.
This requires (at least) $2n$ calls to \proc{Block}, itself potentially recursive.
A similar cost is paid for every equality IC3 must prove. In this example, IC3 proves $r_1 == r_2$ and soon after concludes that the non-interference property holds.
In general, IC3 may need many equalities relating corresponding components in (the two copies of) a self-composition in order to prove non-interference for a large design.


\secmc addresses these inefficiencies through two techniques:
\emph{Symmetric State Exploration} and \emph{Adding Equivalence Predicates}, respectively. 
\cref{alg:block_new} shows the fragment of the \proc{Block} procedure that we modify to incorporate these techniques.
The code snippet corresponds to lines~\ref{ln:relative_induction}-\ref{ln:return_block} in Alg.~\ref{alg:block}, omitting the rest of \proc{Block} as it is unchanged.
After IC3 generalizes an unreachable cube $c$, \secmc will call an additional generalization procedure \proc{Predicate\_replace} to update $c$ to use equivalence predicates where appropriate (\cref{ln:predicate}). It then calls \proc{Symmetric\_cube} to construct a second cube $c_{sym}$ based on the symmetry of self-composition that is also guaranteed to be unreachable (\cref{ln:symmetry}). Both cubes $c$ and $c_{sym}$ are blocked in the current frame (\cref{ln:strengthen_new}).
We further detail the novel techniques following their program order in \cref{alg:block_new}.

\begin{algorithm}[t]
\caption{\secmc modification on \proc{Block} (marked as blue)}
\label{alg:block_new}
\begin{algorithmic}[1]
\If{$\mathsf{UNSAT}(F_{k-1} \land \lnot s(x) \land Tr \land  s(x'))$}
\Comment{reachability test}
  \State $c \gets$ \Call{Generalize}{$ s, k$}
  \Comment{a generalized cube}
  \State \added{$c \gets$ \Call{Predicate\_replace}{c}}
  \label{ln:predicate}
  \State \added{$c_{sym} \gets $ \Call{Symmetric\_cube}{$c$}}
  \label{ln:symmetry}
  \For{$j \gets 1$ \textbf{to} $k$}
    \State $F_j \gets F_j \land \lnot c \; \added{\land \lnot c_{sym}}$ 
    \label{ln:strengthen_new}
  \EndFor
  \State \Return \textbf{true}
\EndIf
\end{algorithmic}
\end{algorithm}

\subsection{Adding Equivalence Predicates}
\label{sec:design-predicates}

To help IC3 prove equalities (addressing Inefficiency 2), we augment the self-composition with auxiliary boolean variables, called \emph{(in)equivalence predicates}, that IC3 can block as part of a cube in order to guarantee a pair of corresponding registers between the two copies are equivalent. 
(We regret the double-negative, IC3 \underline{blocks} \underline{in}equivalence predicates.)
We detail our augmentation and the \proc{Predicate\_replace} procedure that uses these predicates to further generalize a cube.

\begin{figure}[t]
    \centering
    \vspace{-1em}
    \includegraphics[width=0.35\columnwidth]{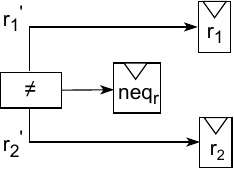}
    \caption{Equivalence Predicate Construction}
    \vspace{-1.5em}
    \label{fig:equiv_predicate}
\end{figure}

\subsubsection{Introducing Word-Level Inequivalence Predicates}
For every pair of registers $r_1, r_2$ between the two copies in a self-composition, a 1-bit register, $neq_r$, is added to represent whether these two registers are unequal. 
We update the self-composition as shown in~\cref{fig:equiv_predicate}, adding a word-level not-equals comparison between the next state values of $r_1$ and $r_2$, and store the result in $neq_r$.
If IC3 blocks $neq_r$, then it has proven (in a single cube!) that the two registers $r_1, r_2$ are always equal, regardless of the value of each copy's secret.

\subsubsection{Predicate Replacement in the Generalized Cube}
Given an augmented self-composition, IC3 is fully capable of reasoning about inequivalence predicates out-of-the-box. However, we found that in practice, many opportunities to block an inequivalence predicate were missed by the standard IC3 generalization procedure.
\secmc adds an additional \proc{Predicate\_replace} step (\cref{alg:block_new}, \cref{ln:predicate}) on each unreachable cube $c$ after generalization, seeking to smartly replace registers in $c$ with their corresponding $neq$ inequivalence predicates while still preserving unreachability.
Continuing our running example (\cref{fig:motivating_example}), 
if IC3 finds an unreachable, generalized cube $c:= (r_1[0]\land \lnot r_2[0])$, it may be prudent to replace it with $(neq_r)$, rather than waiting to block the other half of $r_1[0] \neq r_2[0]$, and then incrementally blocking each bit of $r$ ($2n$ calls to \proc{Block}).
We are careful, as part of this replacement, to use additional reachability tests (\cref{alg:block}, \cref{ln:relative_induction}) to ensure the new cube remains unreachable.

In \proc{Predicate\_replace}, we use the pattern $x_a[i] \land \lnot x_b[i]$ (for some register $x$; copies $a, b$; bit $i$) to identify opportunities to use an inequivalence predicate $neq_x$, as the pattern is a common part of cubes that facilitate equality proofs in IC3.
Given a cube, we call each subformula that matches this pattern an \emph{inequivalence group}, and choose whether or not to replace it with its corresponding predicate heuristically.
We discuss three representative heuristics below, and compare their performance in our evaluation~\cref{sec:evaluation}.

\begin{figure}
    \includegraphics[width=\columnwidth]{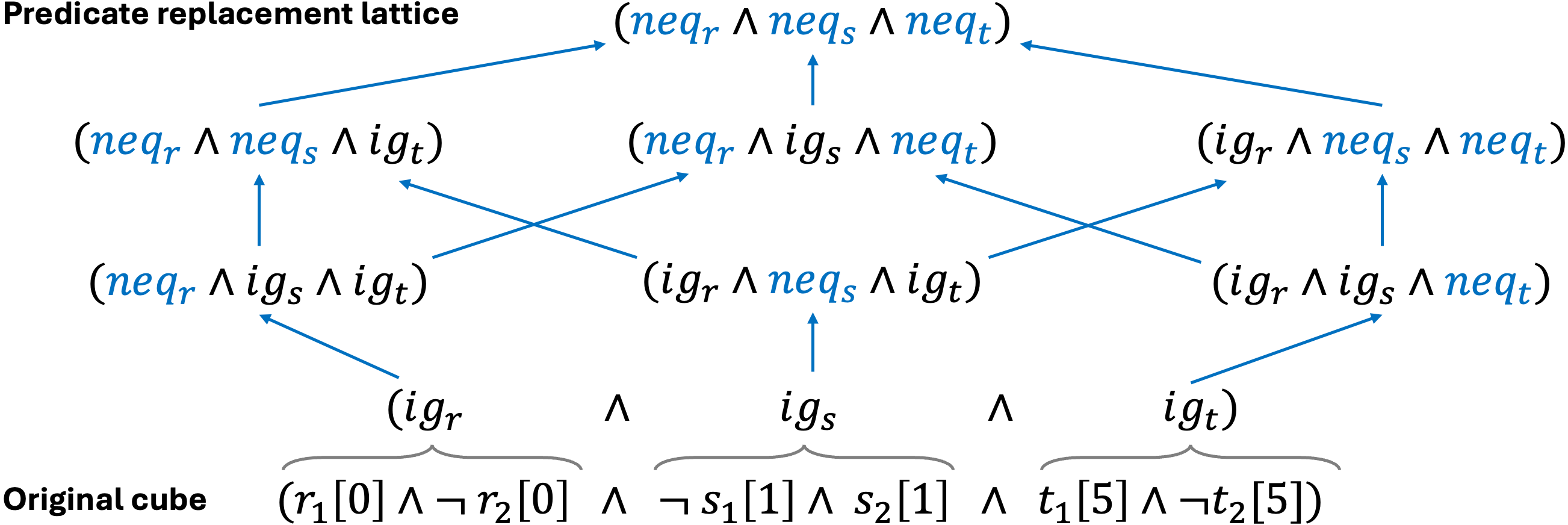}
    \caption{A lattice (partial order) of possible predicate replacements given the cube at the bottom of the lattice.}
    \vspace{-1.6em}
    \label{fig:heuristic_lattice}
\end{figure}

\textit{\textbf{All-or-Nothing Replacement}} replaces all inequivalence groups with their corresponding $neq$ variables and does one reachability test. 
For example, the cube at the bottom of \cref{fig:heuristic_lattice}, would be generalized by \emph{All-or-Nothing Replacement} into the cube at top of the lattice. \proc{Predicate\_replace} would check its reachability, returning the top if it is unreachable, and otherwise the original cube at the bottom.

\textit{\textbf{Maximal Replacement}} cumulatively replaces the inequivalence groups one by one, skipping groups if the replacement causes the cube to be reachable.
In \cref{fig:heuristic_lattice}, \emph{Maximal Replacement} would traverse from the bottom of the lattice, moving up when the cube is unreachable and to the right (another parent of the last unreachable cube) when the cube is not.
\proc{Predicate\_replace} returns the last unreachable cube, defaulting to the original, at the bottom.
Note that different orders of inequivalence groups can result in different final cubes, but each is optimal in that no additional inequivalence group can be replaced without becoming reachable.

\textit{\textbf{Maximum Replacement}} identifies the set of all cubes that \emph{could} be returned by Maximal Replacement, \emph{across all orderings} of inequivalence groups.
To do this efficiently, \emph{Maximum Replacement} would traverse \cref{fig:heuristic_lattice} by starting with a singleton set $M$, containing the cube at the top of the lattice. For each cube added to $M$, we perform a reachability test. Unreachable cubes are left as-is, but reachable cubes are removed from the $M$ and replaced with their lower-cover (all direct children in the lattice), filtering out cubes already subsumed by a cube in $M$.
\proc{Predicate\_replace} returns the final cubes in set $M$, so that they can all be blocked in \cref{alg:block_new}.

These heuristics represent a tradeoff between using fewer SAT queries or blocking more states. 
For a cube with $m$ inequivalence groups:
\begin{itemize*}[label={\hspace{-0.5em}}]
    \item \textit{All-or-Nothing Replacement} costs one query, but gives up immediately;
    \item \textit{Maximal Replacement} costs $m$ queries, but is affected by the order in which inequivalence groups are replaced;
    \item \textit{Maximum Replacement} costs up to $2^m$ queries (in the worst case), but exhaustively blocks as much of the unreachable state-space as possible.
\end{itemize*}

\subsection{Symmetric State Exploration}
\label{sec:design-symmetry}

Given any reachability property of a self-composition (e.g., that a given cube is unreachable), it is always possible to construct a symmetric reachability property that swaps variables of the two copies and continues to hold on the same self-composition.
To see this, note the symmetry of the space of reachable states: 
Suppose a state $(copy_1, copy_2)$ can be reached with input $(pub, sec_1, sec_2)$, where $pub$ corresponds to public inputs and $sec_1, sec_2$ correspond to the secret inputs of each copy (over multiple cycles).
Then the symmetric state $(copy_2, copy_1)$ must be reachable through input that swaps the secrets of each copy, $(pub, sec_2, sec_1)$, because the two copies of the self-composition are identical.

In \secmc, when we construct the self-composition, we record pairs of corresponding state variables (registers) between the two copies. During verification, whenever a cube is proved unreachable, we construct the symmetric unreachable cube by substituting variables with their paired counterpart. The one subtlety is that after predicate replacement, unreachable cubes may contain some equivalence predicates, which are auxiliary variables not part of either copy of the self-composition. Because these predicates check equality they are their own symmetric counterpart and thus are retained as is in the symmetric unreachable cube. 





\section{Evaluation}
\label{sec:evaluation}

We evaluate the performance of our \secmc implementation against the two open-source IC3 implementations it modifies, ABC-PDR~\cite{brayton2010abc} and rIC3~\cite{su2025ric3}.
As part of our evaluation, we collected and organized several open-source hardware designs into a benchmark for information flow checking in hardware. We contribute this benchmark to the community and briefly overview its contents here.
The implementation and evaluation code are open-sourced (https://github.com/qinhant/SecIC3).

\subsection{Implementation}


\secmc is implemented as a Python frontend connected to custom versions of ABC-PDR (C) and rIC3 (Rust).
Given a self-composition circuit and a non-interference property, the frontend (i) (optionally) adds inequivalence predicates, and (ii) uses the Yosys open synthesis suite~\cite{wolf2016yosys} to convert the circuit into an And-Inverter-Graph (AIG) for verification.
%
This AIG is verified by ABC-PDR or rIC3, which we have customized with additional options to enable our two verification techniques:
\begin{itemize}
\item \textit{Symmetry} enables symmetric state exploration (\S~\ref{sec:design-symmetry})
\item \textit{All-or-nothing/Maximal/Maximum} enables predicate replacement following the three heuristics in \S~\ref{sec:design-predicates}.
\end{itemize}

As the two techniques are independent, we will evaluate all configurations of symmetry and predicate replacement.

\subsection{Hardware Non-Interference Checking Benchmark}

\begin{table}[t]

\centering
\small
\begin{tblr}{
  width=\linewidth,
  colspec={|c|X[3,l,m]|},
  row{1-Z} = {valign=m},
  hlines
}
\SetCell{valign=m} Multiplier &
\SetCell{valign=m}
{64-bit shift-and-add integer multiplier \\
194 inputs, 530 registers, 7926 gates
 } \\
\SetCell{valign=m} Modexp~\cite{website:rsa_verilog} & 
\SetCell{valign=m}
{64-bit modular exponentiation calculator \\
258 inputs, 132 registers, 1300 gates
} \\
\SetCell{valign=m} GCD~\cite{website:GCD_verilog} & 
\SetCell{valign=m}
{8-bit greatest common divisor calculator\\
27 inputs, 38 registers, 836 gates
} \\
\SetCell{valign=m} FP\_ADD~\cite{website:fpu_verilog} & 
\SetCell{valign=m}
{Single-precision floating pointer adder\\
197 inputs, 488 registers, 8695 gates
} \\
\SetCell{valign=m} FP\_MUL~\cite{website:fpu_verilog} & 
\SetCell{valign=m}
{Single-precision floating pointer multiplier\\
197 inputs, 512 registers, 18128 gates
} \\
\SetCell{valign=m} FP\_DIV~\cite{website:fpu_verilog} & 
\SetCell{valign=m}
{Single-precision floating pointer divider\\
197 inputs, 836 registers, 9531 gates
} \\
\SetCell{valign=m} SecEnclave~\cite{website:se_prototype} & 
\SetCell{valign=m}
{Secure enclave with AES encryption/decryption and integer arithmetic\\
524 inputs, 2866 registers, 1199964 gates
} \\
\SetCell{valign=m} Cache~\cite{website:cache_verilog} & 
\SetCell{valign=m} 
{4-way 4-word cache\\
270 inputs, 718 registers, 5630 gates
} \\
\SetCell{valign=m} Sodor~\cite{sodor} & 
\SetCell{valign=m} 
{5-stage in-order RISCV core (RV32I)\\
132 inputs, 4004 registers, 51063 gates
} \\
\SetCell{valign=m} Rocket~\cite{rocket} & 
\SetCell{valign=m} 
{5-stage in-order RISCV core (RV64GC)\\
521 inputs, 9026 registers, 123774 gates
} \\
\end{tblr}
\vspace{-0.5em}
\caption{Non-Interference Checking Benchmark}
\vspace{-2em}
\label{tab:benchmark}
\end{table}

We address the lack of a standard benchmark for evaluating information flow in hardware designs through constructing a benchmark with 10 representative open-source hardware designs including arithmetic modules, memory modules and processor cores, as shown in \cref{tab:benchmark}.
The statistics of the self-composition circuits are also shown in the table.
The only output of these circuits is the property output.

For all designs, the non-interference property we are proving models \emph{constant-time execution}, a form of confidentiality that is key in preventing microarchitectural timing side channels.
For arithmetic modules, the source is the input operands and the sink is the {\small{\texttt{out\_valid}}} signal, which serves as a proxy for execution time.
For the cache module, the source is the input address and the sink is the output {\small{\texttt{hit}}} signal.
For two processor cores, the source is the register file and the sink is the instruction {\small{\texttt{retire}}} signal.
The register file is initialized with a free secret variable and operates normally after initialization.
All 10 designs are common in that they do not satisfy constant-time execution when all inputs are free, but they are constant-time with constraints on the inputs.
This is consistent with real practice.
For example, the Sodor processor has constant-time and non-constant-time instructions, and we want a security proof for the constant-time instructions.
Thus, we constrain the instruction memory to have only R-type and I-type instructions, with no branch instructions allowed.

\subsection{Experiment Setup}
We use the default IC3 mode in ABC-PDR and rIC3 as two baselines and evaluate \secmc on each of them with 7 configurations: \textit{Symmetry}, \textit{All-or-Nothing}, \textit{Maximal}, \textit{Maximum}, \textit{Symmetry+All-or-Nothing}, \textit{Symmetry+Maximal}, \textit{Symmetry+Maximum}.
The time limit for each model checking task is 1 hour.
Experiments are run on a machine with 10 M1-MAX cores (8 3.2Hz cores and 2 2.1Hz cores) and 32GB memory.

\subsection{Experiment Results}
\begin{table*}[t]

\centering
\small
\begin{tblr}{
  width=\linewidth,
  colspec={|c|c|c|c|c|c|c|c|c|},
  row{1-Z} = {valign=m},
  hlines
}
\SetCell{valign=m}  & ABC-PDR & Sym & AoN & Maximal & Maximum & Sym+AoN & Sym+Maximal & Sym+Maximum \\
\SetCell{valign=m} Multiplier & 11.5 & 3.0& 0.1& 0.1& 0.1& 0.1& 0.1& 0.1 \\
\SetCell{valign=m} Modexp & 1.5 & 0.6& 1.8& 1.6& 0.7& 1.6& 1.5& 0.4 \\
\SetCell{valign=m} GCD & 74.3 & 75.1& 12.7& 39.3& 9.3& 35.1& 131.9& 50.7 \\
\SetCell{valign=m} FP\_ADD & TO(12) & TO(11) & 466.2& TO(11)& 150.3& 143.7& 186.0& TO(13) \\
\SetCell{valign=m} FP\_MUL & TO(12) & TO(9) & TO(11) & TO(13) & TO(12) & TO(12) & TO(12) & TO(13) \\
\SetCell{valign=m} FP\_DIV & TO(61) & TO(59) & TO(57) & TO(52) & TO(43) & TO(59) & TO(65) & TO(36) \\
\SetCell{valign=m} SecEnclave & TO(7) & 608.4 & 2.5 & 3.6 & 1.9 & 2.4  & 2.1 & 1.9 \\
\SetCell{valign=m} Cache & 1419.1 & 370.5 & 783.0 & TO(18) & TO(15) & 1112.5 & 162.0 & 220.6 \\
\SetCell{valign=m} Sodor & 5.0  & 2.7 & 0.9 & 0.8 & 0.9 & 0.4 & 0.4 & 0.4 \\
\SetCell{valign=m} Rocket & 4.0 & 2.3 & 3.5 & 2.8 & 2.7 & 1.9 & 1.8 & 1.8 \\
\end{tblr}
\vspace{-0.25em}
\caption{Evaluation results on ABC-PDR. Sym stands for Symmetry and AoN stands for All-or-Nothing. 
Time is shown in seconds. For timeout cases (TO), the proof bound is shown in ().}
\vspace{-1em}
\label{tab:result_abc}
\end{table*}

\begin{table*}[t]

\centering
\small
\begin{tblr}{
  width=\linewidth,
  colspec={|c|c|c|c|c|c|c|c|c|},
  row{1-Z} = {valign=m},
  hlines
}
\SetCell{valign=m}  & rIC3 & Sym & AoN & Maximal & Maximum & Sym+AoN & Sym+Maximal & Sym+Maximum \\
\SetCell{valign=m} Multiplier & 2.7 & 2.7 & 0.2 & 0.2 & 0.3 & 0.2 & 0.2 & 0.3 \\
\SetCell{valign=m} Modexp & 0.6 & 0.3 & 0.7 & 0.8 & 1.9 & 0.6 & 1.0 & 1.7 \\
\SetCell{valign=m} GCD & 7.1 & 13.5 & 0.8 & 3.3 & 26.2 & 3.7 & 1.4 & 2.2 \\
\SetCell{valign=m} FP\_ADD & 249.0 & 576.3 & 1925.4 & 62.4 & 80.5 & 44.8 & 69.0 & 114.5 \\
\SetCell{valign=m} FP\_MUL & TO(14) & TO(14) & TO(14) & TO(14) & TO(14) & TO(14) & TO(15) & TO(15) \\
\SetCell{valign=m} FP\_DIV & TO(62) & TO(77) & TO(103) & TO(60) & TO(60) & TO(103) & TO(67) & TO(56) \\
\SetCell{valign=m} SecEnclave & 92.1 & 97.7 & 457.5 & 496.0 & 764.5 & 98.7 & 93.4 & 130.5 \\
\SetCell{valign=m} Cache & 2970.2 & 44.0 & 207.3 & 71.4 & 94.5 & 112.1 & 87.2 & 88.4 \\
\SetCell{valign=m} Sodor & 1.5 & 1.9 & 2.0 & 2.1 & 3.3 & 2.0 & 1.9 & 4.0 \\
\SetCell{valign=m} Rocket & 6.3 & 6.5 & 6.9 & 7.5 & 9.3 & 7.3 & 8.4 & 12.0 \\
\end{tblr}

\caption{Evaluation results on rIC3. Headings, units, and notation match \cref{tab:result_abc}.
}
\vspace{-2em}
\label{tab:result_ric3}
\end{table*}

Comparisons with ABC-PDR are shown in~\cref{tab:result_abc}. Relative to the baseline, \secmc can prove non-interference for two more designs, FP\_ADD and SecEnclave.
For others, \secmc significantly improves the proof time or proof bounds (A proof bound $n$ shows the property holds in the first $n$ cycles.).

Comparisons with rIC3 are shown in \cref{tab:result_ric3}. While \secmc does not verify more designs than baseline, it does improve the proof bound of FP\_DIV from 62 to 103, and the proof bound of FP\_MUL from 14 to 15.
For other cases, \eg Cache, \secmc significantly improves proof time.

Among different configurations of \secmc, no option is dominant over others in either table. However, we note that for some designs (and solvers), such as Cache in \cref{tab:result_abc} (ABC-PDR) or SecEnclave in \cref{tab:result_ric3} (rIC3), using predicate replacement alone can significantly deteroriate performance relative to baseline.
Therefore, \textit{Symmetry} along with predicate replacement should be the default configuration of \secmc.

\begin{figure}[t]
    \centering
    \includegraphics[width=\columnwidth,
    trim=0 0 0 1.1em,
    clip]{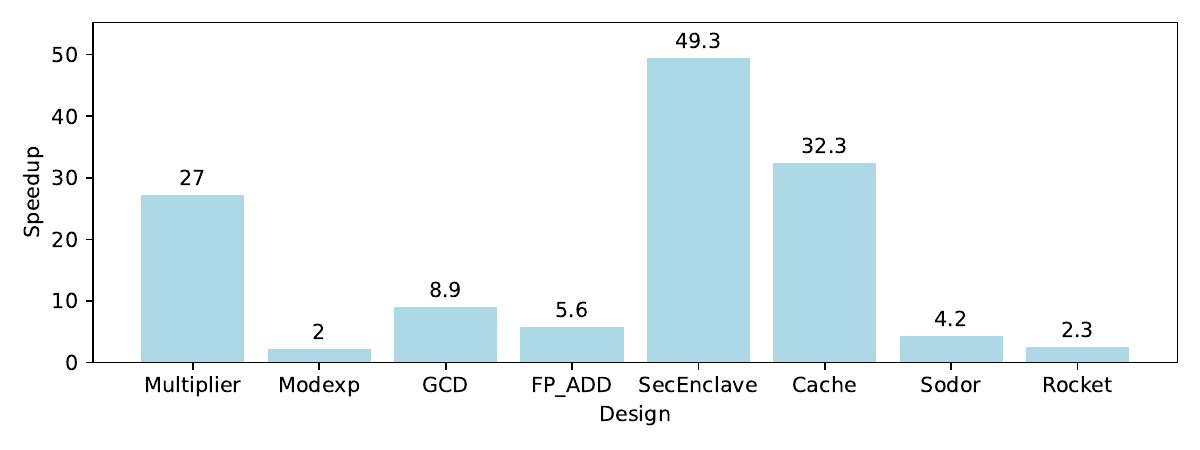}
    \vspace{-2em}
    \caption{Proof Speedup Statistics}
    \vspace{-2em}
    \label{fig:speedup}
\end{figure}

\cref{fig:speedup} shows speedup statistics for each successfully verified design, dividing the best result among the two baselines by the best result among all \secmc configurations (across both backends).
\secmc significantly improves the proof efficiency for all designs in the benchmark, with speedups ranging from 2.0x to 49.3x.
The average speedup is 16.5x among these 8 cases.
For the timeout cases FP\_MUL and FP\_DIV (not shown in~\cref{fig:speedup}), \secmc can improve the proof bound to 15 and 103 cycles, respectively, while the best among the two baselines can only prove 14 and 62 cycles.

\section{Related Work}
\label{sec:related}
\subsection{Self-Composition}
Self-composition was first proposed by Barthe et al.~\cite{barthe2004secure} to check program security properties.
Many subsequent efforts focused on effective construction of self-composed programs (e.g.,~\cite{TerauchiSas05,BartheFM11,shemer2019property}) to improve verification performance, but they largely used model checkers as black boxes during safety verification of the self-composition.

Self-composition has been widely used in verifying hardware security properties such as constant-time execution~\cite{IODINE,v2021solver} and speculative leakage contracts~\cite{tan2025rtl, jauch2023secure, wang2023specification}.
These properties are all based on non-interference but they differ in their assumptions.
For constant-time execution properties (as in our benchmarks), assumptions are typically logical expressions on input variables.
In contrast, for speculative leakage contracts, the assumptions can themselves be non-interference properties --- we hope to address solver customization for such properties in future work.

\subsection{Hardware Information Flow Tracking}

Aside from self-composition, \emph{information flow tracking} has also been used to prove non-interference in hardware~\cite{glift,cellift,ardeshiricham2017register}. In this technique, the design is extended with additional ``taint'' signals that track if a wire/register either (i) is a secret or (ii) may depend on secrets based on the taint signals of its fan-in. This \emph{taint propagation} is conservative~\cite{hu2011theoretical}: proving that the taint signals of public outputs are always low guarantees non-interference, but counterexamples where output taint signals are high may be spurious (i.e., they do not leak secrets).
Thus, while information flow tracking provides an alternative trade-off point in checking information flow, self-composition remains the most precise way to do this.


\subsection{IC3 Variants}
Since the proposal of IC3 in 2011~\cite{bradley2011sat}, various variants have emerged.
Word-level IC3 such as IC3ia~\cite{cimatti2016infinite} and IC3sa~\cite{goel2019model} intends to create word-level abstraction of hardware design and reason about it in first-order logic using SMT solvers.
This avoids the expensive bit-blasting in bit-level reasoning, but additionally requires abstraction refinement, which may be more expensive.
The equivalence predicates in \secmc also add some word-level reasoning ability to IC3, but it does not introduce abstraction and thus avoids abstraction refinement.

PDRER~\cite{luka2025property} dynamically adds auxiliary variables to the circuit during model checking based on the lemmas IC3 has learned so far.
Those variables can then be used in IC3 to speed up verification.
This idea is similar to the predicate variables in \secmc, but the equivalence predicates in \secmc are added based on the knowledge of the self-composition design rather than the patterns in the learned lemmas.

\section{Conclusion}
In this paper, we propose \secmc, a customized hardware model checker for hardware security verification based on the IC3 model checking algorithm.
\secmc leverages the special structure of the self-composition circuit used in proving the non-interference property to extend IC3 with two key techniques, \textit{symmetric state exploration} and \textit{adding equivalence predicates} to guide the model checker.
We implemented \secmc on top of (i) ABC-PDR, a widely used open-source IC3 implementation and (ii) rIC3, the winner of the 2025 Hardware Model Checking Competition in both the bit-level and word-level tracks, and evaluated its performance on a new security verification benchmark developed as part of this project consisting of 10 hardware designs.
The evaluation results have shown that \secmc significantly improves the proof performance in security verification and can bring up to 49.3x proof speedup or increase the proof bound within 1 hour by at most 41 cycles compared to state-of-the-art baseline implementations.



\let\section\spacysection
\let\subsection\spacysubsection
\let\subsubsection\spacysubsubsection

\newpage
\bibliographystyle{IEEEtran}
\bibliography{ref,ref-ag}

\end{document}